\documentclass[aps,superscriptaddress,notitlepage]{revtex4-1}
\usepackage{graphicx}
\usepackage{epstopdf}
\usepackage{amsmath}
\usepackage{bm}
\usepackage{times}
\usepackage{lineno}
\usepackage{color}
\usepackage[utf8]{inputenc}
\usepackage{booktabs}

\newcommand{\ems}{\sqrt{s_{NN}}}
\newcommand{\pt}{p_{\rm T}}
\newcommand{\Dr}{\Delta R}
\newcommand{\Dp}{\Delta P}

\begin{document}
\title{Light nuclei elliptic flow at mid-rapidity in $\sqrt{s_{NN}} = 3.0-3.9$ GeV Au+Au collisions using coalescence model}

\author{Y. Xu}
\affiliation{School of Physics, Harbin Institute of Technology, Harbin 150001, China}

\author{X. H. He}
\affiliation{Institute of Modern Physics, Chinese Academy of Sciences, Lanzhou 730000, China}
\affiliation{School of Nuclear Science and Technology, University of Chinese Academy of Sciences, Beijing 100049, China}
\affiliation{Heavy lon Science and Technology Key Laboratory, Institute of Modern Physics, Chinese Academy of Sciences, Lanzhou 730000, China}

\author{Y. P. Zhang}
\affiliation{Institute of Modern Physics, Chinese Academy of Sciences, Lanzhou 730000, China}
\affiliation{School of Nuclear Science and Technology, University of Chinese Academy of Sciences, Beijing 100049, China}
\affiliation{Heavy lon Science and Technology Key Laboratory, Institute of Modern Physics, Chinese Academy of Sciences, Lanzhou 730000, China}


\begin{abstract}
Light nuclei collective flow is an important probe for understanding their production mechanisms in heavy-ion collisions. 
The STAR collaboration has reported that the atomic mass number ($A$) scaling of light nuclei elliptic flow $v_2$ is broken at $\sqrt{s_{NN}} = 3.0-3.9$ GeV. 
The observations reveals that, while protons maintain negative $v_2$ values at mid-rapidity at both 3.0 and 3.2 GeV,
light nuclei $v_2$ exhibit a sign change from negative at 3.0 GeV to positive at 3.2 GeV.
In this study, we investigate $v_2$ of protons and deuterons in mid-central Au+Au Collisions at $\sqrt{s_{NN}} =$ 3.0, 3.2, 3.5 and 3.9 GeV using the JAM2 microscopic transport model.
Deuterons are formed via nucleon coalescence, with the spatial distance ${\Delta R}$ 
and momentum difference ${\Delta P}$ between constituent protons and neutrons serving as the coalescence criteria.
Our calculations successfully reproduce the sign change in deuteron $v_2$ at 3.2 GeV. We observe a strong dependence of nucleon coalescence probability on the azimuthal angle relative to the reaction plane. This effect is primarily driven by the transverse momentum dependence of the mean spatial $\langle {\Delta R} \rangle$ 
and momentum $\langle {\Delta P} \rangle$ separations between nucleon pairs, which vary with the nucleon azimuthal angle.  Moreover, our analysis demonstrates that the stiffness of the nuclear equation of state plays a crucial role in determining the energy dependence of this sign change in deuteron $v_2$ at $\sqrt{s_{NN}}=3.2$ GeV.

\end{abstract}

\maketitle

\section{Introduction}
The study of nuclear matter at extreme temperatures and densities offers valuable  insights into the properties of strongly interacting many-body systems described by quantum chromodynamics (QCD). The macroscopic properties of nuclear matter under extreme conditions are most evident in measurable collective features, which represent the common dynamics of multiple particles produced in a single reaction. These collective features are manifested as collective flow, characterized by the motion of numerous outgoing particles exhibiting either aligned directional movement or uniform velocity magnitudes~\cite{annurev:/content/journals/10.1146/annurev.nucl.49.1.581}. The observation of hadron elliptic flow ($v_2$) shows approximate number-of-constituent-quark (NCQ) scaling at large $\pt$ at high beam energies at RHIC and LHC. This scaling is interpreted as a signature for emergence of the quark-gluon plasma (QGP) formed during these collisions~\cite{VOLOSHIN2003379c,PhysRevLett.91.092301,PhysRevLett.95.122301,DONG2004328,PhysRevC.85.064914,ALICE:2014wao}. 

In heavy-ion collisions, the production mechanism of light nuclei (such as deuteron, triton, $^{3}$He, $^{4}$He, etc.) remains a topic of ongoing debate.
A widely accepted theoretical model is that these nuclei are formed through the coalescence of nucleons, which can either newly produced or transported from the colliding nuclei~\cite{PhysRevLett.37.667,liu2024light,zhu2015light, zhao2018spectra,kachelriess2023effect}.
According to this model, light nuclei emerge at the late stage of the collision evolution when the constituent nucleons come into close in both the coordinate space and momentum space~\cite{SATO1981153, PhysRevC.99.014901}. 
A key features of this model, analogous to the NCQ scaling in the flow of hadrons, is that the collective flow of light nuclei is expected to exhibit scaling behavior
with respect to their atomic mass number ($A$). 
While the distinction between quark coalescence and nucleon coalescence lies in the fact that, in nucleon coalescence, both the momentum and spatial distribution of the constituent nucleons$-$protons$-$can be directly measured in heavy-ion collision experiments, along with the resulting light nuclei.

The STAR collaboration has reported that the $A$ scaling of light nuclei $v_2$ holds in low transverse momentum range ( $\pt/A < 1.5 \rm{GeV}/c$) at $\ems = 7.7-200$ GeV~\cite{PhysRevC.94.034908}.
The results of the transport plus coalescence model for light nuclei $v_2$ are also consistent with the experimental measurements. 
While the recent measurements at $\ems = 3.0-3.9$ GeV Au+Au collisions by the STAR fixed-target experiments have revealed distinct scaling patterns for light nuclei anisotropic flow~\cite{2022136941,refId0,flowqm2023}. 
The directed flow ($v_1$) of light nuclei clearly follows an $A$ scaling, strongly supporting coalescence as the dominant production mechanism for these clusters.
In contrast, their $v_2$ shows breakdown of the $A$ scaling behavior under the same collision conditions. 
Moreover, the proton $v_2$ values are negative in mid-rapidity for collision energies of 3.0 GeV and 3.2 GeV, transitioning to positive values above 3.2 GeV~\cite{flowqm2023}. 
Similarly, for deuteron and $^{3}$He, the $v_2$ values are negative at mid-rapidity at 3.0 GeV but approach zero and become positive 3.2 GeV, respectively.

The negative $v_2$ at low energies is attributed to the shadowing effect of the spectators in the collision. 
Given that the transition energy for the sign change of $v_2$ differs between protons and light nuclei, essential to investigate whether this shadowing effect exhibits mass-dependent behavior or if the final phase space distribution of nucleons plays a more dominant role. Such an investigation could provide critical insights into the formation time and mechanisms of light nuclei in heavy-ion collisions.

In this paper, we employ the newly developed Jet AA Microscopic transport model (JAM2)~\cite{PhysRevC.61.024901,PhysRevC.105.014911}, combined with a nucleon coalescence model, to calculate the $v_2$ of protons and deuterons in mid-central Au+Au collisions at $\ems =  3.0-3.9$ GeV. We investigate the nucleon coalescence probability as a function of azimuthal angle with respect to the reaction plane for collisions at $\ems = $3.0, 3.2, 3.5 and 3.9 GeV.
Additionally, we  explore the dependence of proton and deuteron $v_2$ on the stiffness of equation-of-state (EoS) at the studied energies, providing theoretical predictions that can be directly compared with STAR experimental data.

\section{Method}
The calculation begins with event generation for Au+Au collisions using JAM2 at the specified energies.
Within the JAM2 framework, the initial position of incoming nucleons are sampled according to the nuclear density distribution. 
The nuclear collision is determined by summing the contributions of binary hadron-hadron collisions,  based on their closest approach distances.
Particle production is modeled through 
resonance and string excitations, followed by their subsequent decays.
For this analysis, we employ the mean-field model of JAM2 with an incompressibility parameter of $\kappa=$ 210 MeV and 380 MeV. 

The spatial positions and momenta of protons and neutrons  are recorded at a fixed time of 50 fm/$c$ for subsequent coalescence into light nuclei. 
During the afterburner coalescence stage, deuteron formation occurs when the phase-space distance between proton-neutron pair falls below specified thresholds. For each proton-neutron pair, we calculate the relative spatial distance
$\Dr = \left|R_{1}-R_{2}\right|$ and relative momentum distance $\Dp = \left|P_{1}-P_{2}\right|$ in the rest frame of the pair. A deuteron is formed when both $\Dr < 4.5~{\rm fm}$ and $\Dp < 0.3~{\rm GeV}/c$ are simultaneously satisfied.
The choice of parameters $\Dr$ and $\Dp$ is based on the work of~\cite{Xu_2023}, where these coalescence criteria effectively describe the yield of deuterons at $\ems = $ 3.0 GeV. 
Our analysis excludes light nuclei with mass number $A>2$ (which could otherwise reduce the available nucleon pool for deuteron production), this simplification serves our primary objective focus on understanding the $v_2$ sign change between protons and deuterons and facilitates the estimation of $\Dr$ and $\Dp$ for all proton-neutron pairs. The potential overestimation of deuteron yields in this approximation does not affect our investigation
into the $v_2$ sign change. This is because the nuclei yields of $A>2$ is suppressed compared to deuterons, and the sign change phenomenon in coalescence reflects the phase-space distribution of nucleons rather than absolute yields of light nuclei.

In JAM2, the default event-plane angle is set to zero, and consequently, the particle the particle $v_2$ is calculated as $v_2=\langle\cos{2\phi}\rangle$, where $\phi$ represents the particle's azimuthal angle and the angle brackets denote the average over all particles in all events.
In our analysis, collision centrality is determined using the impact parameter $b$, defined as the minimum distance between the centers of the colliding nuclei.
To facilitate direct comparison with STAR experimental data, all presented results are calculated for impact parameters in the range $b=4.3-8.5~{\rm fm}$, corresponding to the 10-40\% centrality interval.

\section{Results and discussion}

\begin{figure}[htbp]
  \centering
  \includegraphics[scale=0.7]{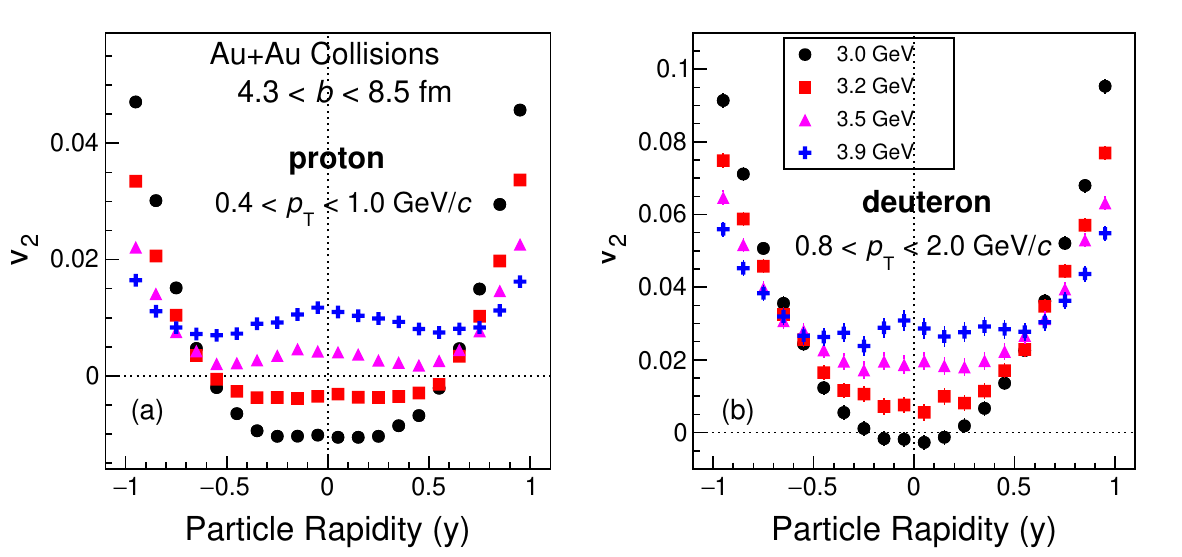}
  \caption{Elliptic flow $v_2$ for (a) protons and (b) deuterons in Au+Au collisions ($b=4.3-8.55~{\rm fm}$) at $\ems =$ 3.0 GeV(solid circles), 3.2 GeV (squares), 3.5 GeV (triangles) and 3.9 GeV (crosses), calculated using JAM2 plus afterburner nucleon coalescence. The results were calculated within $0.4 < p_{\rm{T}} < 1.0$ GeV/$c$ and $0.8 < p_{\rm{T}} < 2.0$ GeV/$c$ for protons and deuterons, respectively.}
  \label{Fig.1}
\end{figure}

Figure~\ref{Fig.1} presents the results of protons and deuterons $v_2$ as a function of particle rapidity in Au+Au collisions at $\ems =$ 3.0, 3.2, 3.5, and 3.9 GeV. The results were calculated using the same transverse momentum range ($0.4 < \pt/A < 1.0$ GeV/$c$) as implemented in the STAR experimental analysis~\cite{flowqm2023}.
At mid-rapidity, the $v_2$ exhibits distinct energy dependent behavior for protons and deuterons. 
Specifically, protons show negative $v_2$ values at $\ems = 3.0, 3.2$ GeV and transitioning to positive values at $\ems = $ 3.5 and 3.9 GeV. In contrast, deuterons show negative $v_2$ values only at $\ems = 3.0$ GeV, maintaining positive values above 3.2 GeV. These results are quantitatively consistent with the experimental measurements from the STAR collaboration~\cite{flowqm2023}.

The negative $v_2$ values at low collision energies can be attributed to the spectator shadowing effect, where the prolonged passage time of non-interacting spectators along the impact parameter direction significantly influences the anisotropic expansion of the fireball.
However, while this effect should in principle affect all particle species uniformly, both experimental data and coalescence model calculations reveal a discrepancy at $\ems = 3.2$ GeV: protons and deuterons have opposite $v_2$ sign at mid-rapidity.
This contradiction arises by the dynamics of deuteron formation. In our calculation, the deuteron are formed through the coalescence of nucleons at a time of 50 fm/$c$, a late stage in the system's evolution when the influence of spectators has become negligible. This delayed formation picture could explain why deuteron $v_2$ maintains positive values even when proton $v_2$ is negative at 3.2 GeV. These results suggested that light nuclei formation in heavy-ion collisions occurs at the late stages, where the spectator effect no longer plays a significant role in shaping the collective flow patterns.

\begin{figure*}[htbp]
  \centering
  \includegraphics[scale=0.84]{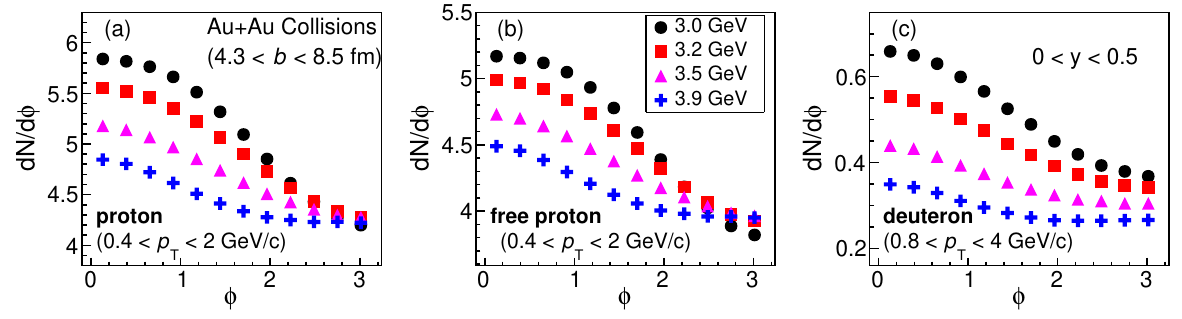}
  \caption{The $dN/d\phi$ distribution of (a) total protons, (b) surviving free protons after coalescence and (c) deuterons in Au+Au collisions ($b=4.3-8.5~{\rm fm}$) at $\ems =$ 3.0 GeV(solid circles), 3.2 GeV (squares), 3.5 GeV (triangles) and 3.9 GeV (crosses). The results are calculated within the rapidity range $0<y<0.5$. All distributions have been normalized by the total number of simulated events. The $v_2$ value was extracted through Fourier expansion using Eq.~(\ref{eq:fourier_expansion}).}
  \label{Fig.2}
\end{figure*}

\begin{figure*}[hbtp]
  \centering
  \includegraphics[scale=0.84]{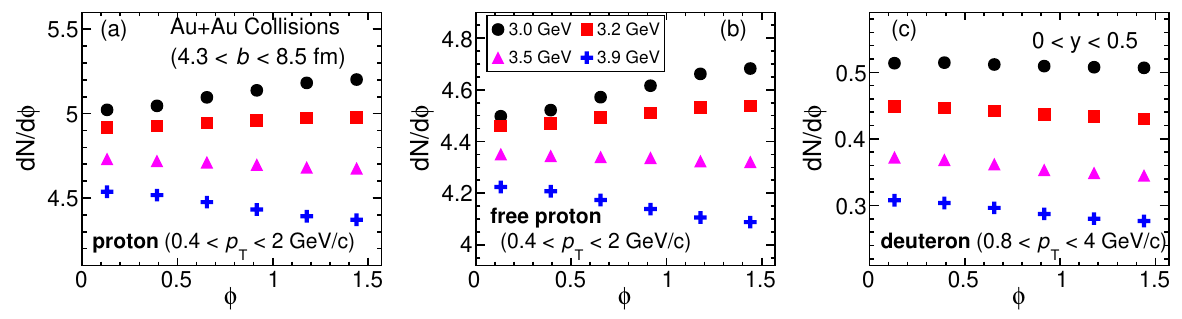}
  \caption{The $dN/d\phi$ distribution folding along $\pi/2$ for (a) total protons, (b) surviving free protons and (c) deuterons in Au+Au collisions ($b=4.3-8.5~{\rm fm}$) at $\ems =$ 3.0 GeV(solid circles), 3.2 GeV (squares), 3.5 GeV (triangles) and 3.9 GeV (crosses). The results are calculated within the rapidity range $0<y<0.5$.}
  \label{Fig.3}
\end{figure*}

To gain deeper quantitative insight into how the coalescence afterburner affects deuteron $v_2$, we examine the azimuthal angle ($\phi$) distributions of particles. The collective flow coefficients $v_n$ is are quantitatively characterized through Fourier expansion of the particle azimuthal distribution with respect to the reaction plane angle, 
which is represented by the $dN/d\phi$ distribution in our calculations. 
In our analysis, we specifically investigate the $dN/d\phi$ distributions to explore the origin of the opposite $v_2$ signs between protons and deuterons at mid-rapidity ($0 < y < 0.5$).

The investigation focuses on three distinct particle populations: (i) protons prior to coalescence, (ii) surviving free protons after coalescence, and (iii) deuterons formed through the coalescence process. Their respective $dN/d\phi$ distributions are shown in Fig.~\ref{Fig.2}. 
We analyze the azimuthal distributions by fitting them with the Fourier expansion:
\begin{equation}
\frac{dN}{d\phi} = \frac{N_0}{2\pi} \left[ 1 + 2v_1\cos\left(\phi\right)+2v_2\cos\left(2\phi\right) \right],
\label{eq:fourier_expansion}
\end{equation}
The $v_2$ values extracted from this fitting show excellent agreement with those presented in Fig.~\ref{Fig.1}, which were obtained via the event plane method using $\langle \cos{(2\phi)} \rangle$. This consistency check is crucial, as the STAR measurements rely on event plane method, while the $dN/d\phi$ distributions provides insights into 
the evolution of $v_2$ throughout the coalescence process.

In the energy range explored in this study, the collision system exhibits a strong $v_1$ alongside nearly vanishing $v_2$ at mid-rapidity. Consequently, the $dN/d\phi$ distribution is dominated by the $v_1$ component.To extract the $v_2$ contribution more clearly, we have applied a folding procedure around $\pi/2$ to the results presented in Fig.~\ref{Fig.2},  with the folded distributions presented in Fig.~\ref{Fig.3}.
These distributions exhibit distinct energy dependencies: 
At $\ems =$ 3.0 GeV, both total protons and surviving free protons after coalescence show pronounced yield enhancements near $\phi=\pi/2$, corresponding to negative $v_2$ values. The distributions become isotropic at $\ems =$ 3.2 and 3.5 GeV, indicating the nearly vanishing $v_2$ for total protons.
While deuterons shows a slight enhancement near $\phi=0$ at 3.2 GeV, corresponding to a positive $v_2$ value.
At the highest investigated energy of $\ems = 3.9$ GeV, the $\phi$ yield show consistent enhancement near $\phi=0$ for all the three particle populations. 
By comparing the $\phi$ distributions before and after coalescence, we can directly examine how the coalescence probability varies with azimuthal angle, providing unique insights into the formation dynamics of light nuclei in heavy-ion collisions.

\begin{figure}[htbp]
  \centering
    \includegraphics[width=0.4\textwidth]{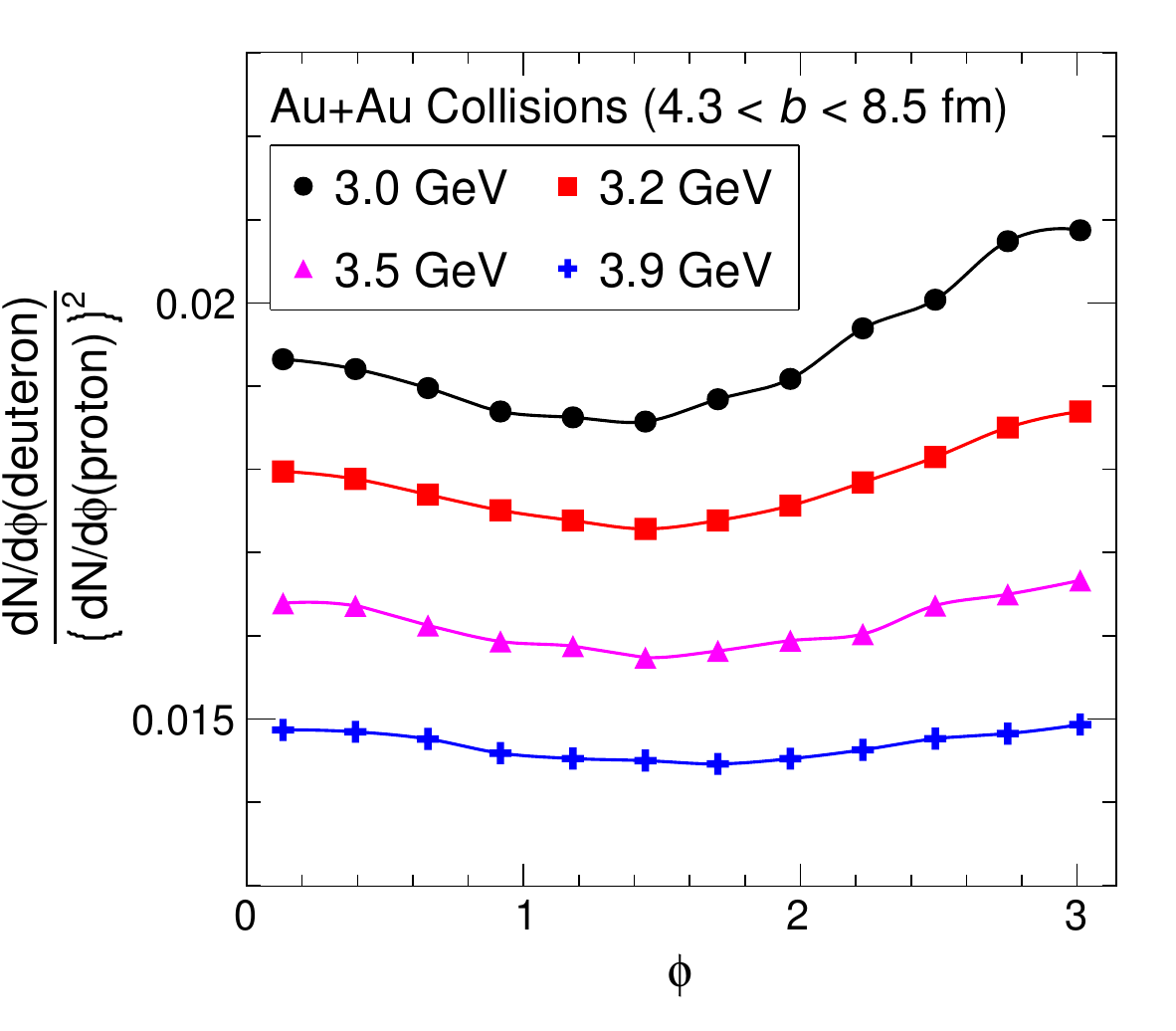}
  \caption{The nucleon coalescence probability as a function of azimuthal angle in Au+Au collisions at $\ems =$ 3.0 GeV(solid circles), 3.2 GeV (squares), 3.5 GeV (triangles) and 3.9 GeV (crosses. The results are calculated within the rapidity range $0<y<0.5$. }
  \label{Fig.4}
\end{figure}

Under the assumption that protons and neutrons have identical  $\phi$ distributions, the coalescence model predicts the deuteron production follows the relation:
\begin{equation}
\frac{dN_{d}}{d\phi}\approx P\left(\frac{dN_{p}}{d\phi}\right)^{2}.
\label{eq2}
\end{equation}
where $P$ represents the coalescence probability. We should not expect the coalescence process to be isotropic across all azimuthal directions, and the coalescence probability $P$ may have  
$\phi$ dependence due to the anisotropic nature of the collision system.
Figure~\ref{Fig.4} shows the coalescence probability $P$ distribution as a function of $\phi$ calculated using the proton and deuteron $\phi$ distributions in Fig.~\ref{Fig.2}. 

The coalescence probability of nucleons shows a clear azimuthal dependence across all studied energies, with  a minimum near $\phi = \pi/2$ (perpendicular to reaction plane direction). While the dependence weakens as the collision energy increases.
This indicates that nucleon coalescence is significantly more probable for particles moving parallel to the reaction plane compared to those moving perpendicularly. Consequently, nucleons contributing to negative $v_2$ around $\phi = \pi/2$ have a reduced probability of coalescing and forming deuterons, thereby explaining the observed sign change in deuteron $v_2$ relative to protons at certain energies. 

\begin{figure}[htbp]
  \centering
  \includegraphics[width=0.7\textwidth]{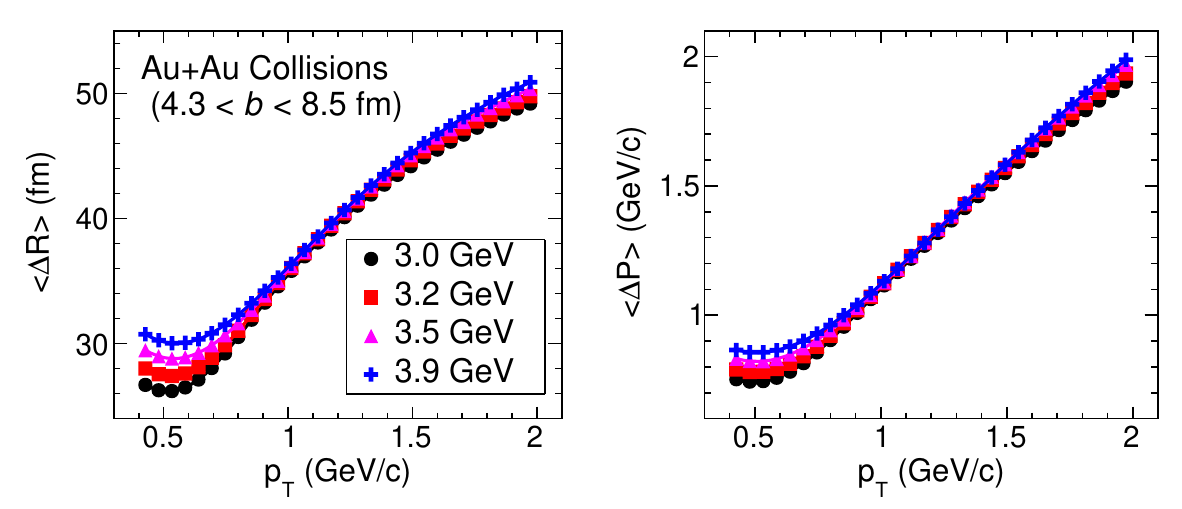}
  \caption{Average spatial ($\langle\Dr\rangle$) and momentum ($\langle\Dp\rangle$) separations of $(p, n)$ pair in Au+Au collisions at $\ems =$ 3.0 GeV(solid circles), 3.2 GeV (squares), 3.5 GeV (triangles) and 3.9 GeV (crosses), calculated using JAM2 plus afterburner nucleon coalescence. The results were calculated within $0<y<0.5$.}
  \label{Fig.5}
\end{figure}

Since the deuteron is formed by satisfying the conditions $\Dr < 4.5~{\rm fm}$ and $\Dp < 0.3~{\rm GeV}/c$,
the minimum coalescence probability near $\phi=\pi/2$ suggests that nucleons oriented perpendicular to the reaction plane are statistically less likely to satisfy the spatial or momentum constraints for deuteron formation compared to those along the reaction plane. To investigate this phenomenon quantitatively, we analyze the average spatial ($\langle\Dr\rangle$) and momentum ($\langle\Dp\rangle$) separations as functions of proton $p_{\rm T}$ in the range of $0.4 < \pt < 2$ GeV/$c$, as shown in Fig.~\ref{Fig.5}. Both $\langle\Dr\rangle$ and $\langle\Dp\rangle$ have strong $\pt$ dependence, increasing monotonically with proton $\pt$ above 0.5 GeV/$c$, where the $\langle\Dr\rangle$ reaches its minimum. The observed systematic increase in both $\langle\Dr\rangle$ and $\langle\Dp\rangle$ with collision energy across the measured $\pt$ range reflects the stronger radial flow and higher pressure gradients developed in the more energetic collisions.

\begin{figure}[htbp]
  \centering
  \includegraphics[width=0.4\textwidth]{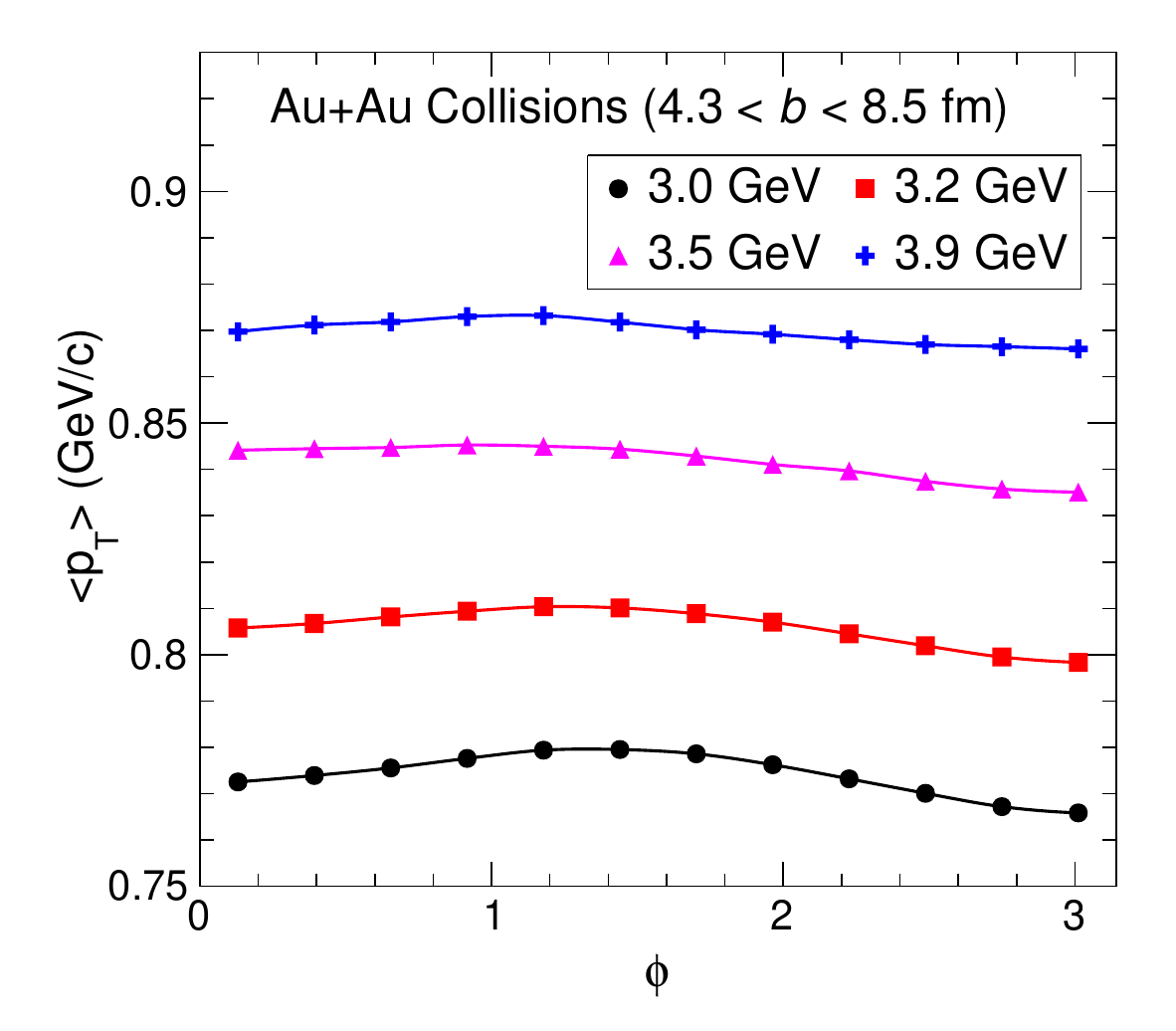}
  \caption{The proton mean $\pt$ distributions as function of $\phi$ in Au+Au collisions at $\ems =$ 3.0 GeV(solid circles), 3.2 GeV (squares), 3.5 GeV (triangles) and 3.9 GeV (crosses). The results were calculated within $0<y<0.5$.}
  \label{Fig.6}
\end{figure}

Figure~\ref{Fig.6} shows the azimuthal angle dependence of mean transverse momentum ($\langle\pt\rangle$) for protons in the range $0.4<\pt<2.0$ GeV/$c$ and $0<y<0.5$  from JAM2 model calculations. A characteristic enhancement of $\langle\pt\rangle$ around $\phi=\pi/2$ is observed at $\ems=3.0-3.5$ GeV, demonstrating the spectator squeeze-out effect at these energies. The stronger transverse momentum boost for protons perpendicular to the reaction plane becomes more pronounced at lower energies due to the increased passage time of spectator matter, which amplifies the anisotropic pressure gradients. By $\ems=4.5$ GeV, the $\langle\pt\rangle$ become nearly isotropic. 

Higher momentum protons around $\phi=\pi/2$ exhibit larger  $\langle\Dr\rangle$ and $\langle\Dp\rangle$ separations (Fig.~\ref{Fig.5}), reducing their coalescence probability (Fig.~\ref{Fig.6}), which directly impacts deuteron formation.
Therefore, this azimuthal dependent $\langle\pt\rangle$ and coalescence probability explains the breaking of mass-number scaling at the studied energies and the sign change of deuteron $v_2$ at $\ems=3.2$ GeV.

\begin{figure}[htbp]
  \centering
  \includegraphics[width=0.4\textwidth]{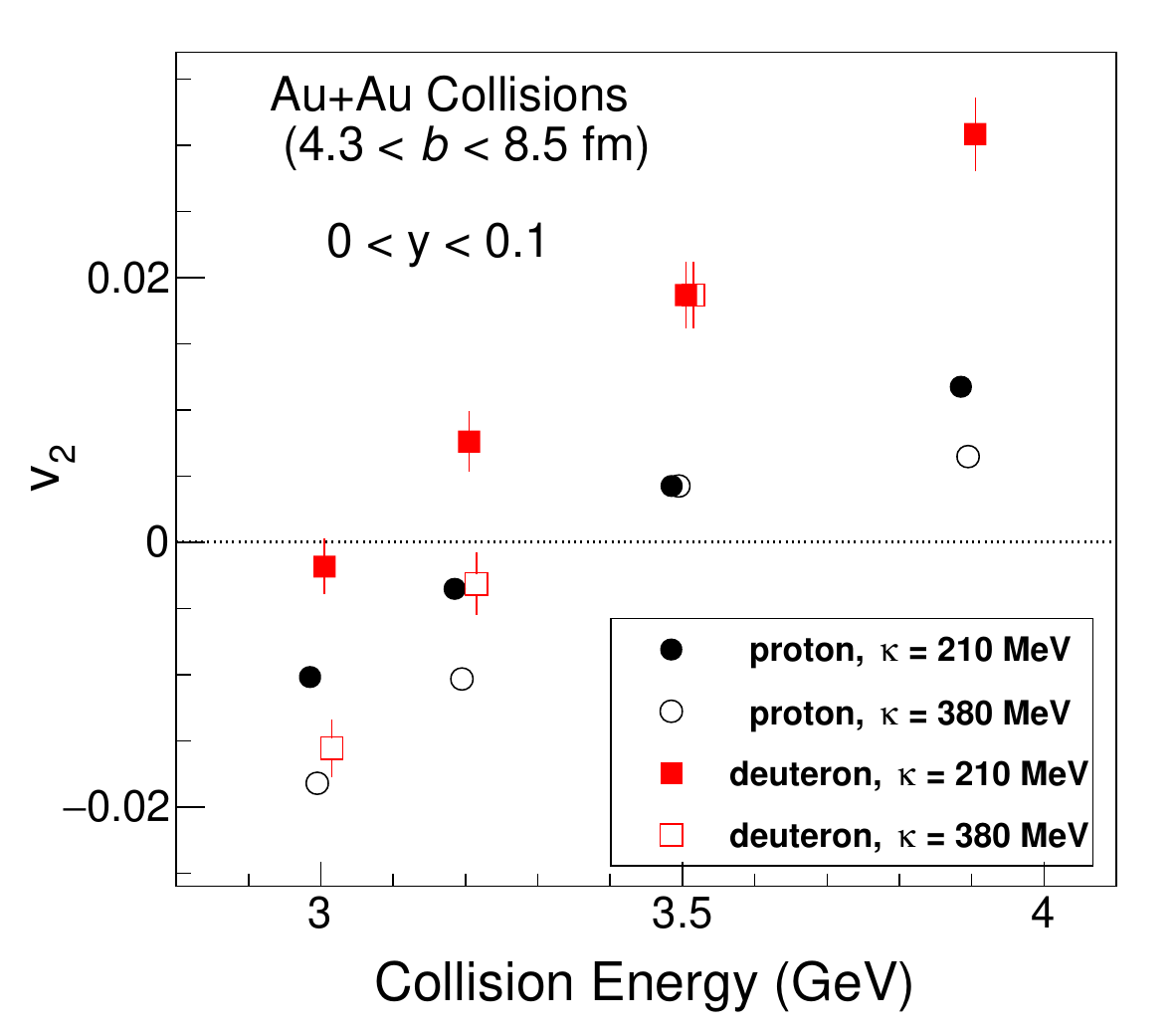}
  \caption{Elliptic flow of protons (circles) and deuterons (squares) at mid-rapidity ($0 < y < 0.1$) in $\ems = $ 3.0, 3.2, 3.5 and 3.9 GeV Au+Au collisions. Results are shown for two nuclear incompressibility parameters:  $\kappa$ = 210 MeV (solid symbols) and 380 MeV (open symbols), representing the soft and hard EoS respectively. }
  \label{Fig.7}
\end{figure}

Figure~\ref{Fig.7} compares the $v_2$ results obtained using two different nuclear incompressibility parameters ($\kappa$ = 210 MeV and 380 MeV), representing soft and stiff EoS respectively. ($\kappa$ = 210 MeV and 380 MeV). At $\ems = 3.0$ GeV, both proton and deuteron $v_2$ value remain negative regardless of the EoS stiffness. But for $\ems =$ 3.5 and 3.9 GeV, all calculated $v_2$ value are positive. The transitional energy of $\ems = 3.2$ GeV shows particularly interesting behavior: While proton $v_2$ stays negative, deuteron $v_2$ shows strong sensitivity to the EoS stiffness, remaining negative for the stiff EoS but turning positive for the soft EoS. 
This striking difference implies that the nuclear EoS plays a crucial role in determining the sign inversion of light nuclei $v_2$. 
Furthermore, the energy dependence of the $v_2$ splitting between protons and deuterons are also sensitive to the EoS stiffness. As the collision energy decrease from $\ems = 3.9$ to 3.0 GeV, the magnitude of this $v_2$ splitting systematically diminishes, with this trend being particularly pronounced for the stiff EoS. At the lowest studied energy ($\ems = 3.0$ GeV), the proton and deuteron $v_2$ become identical for the stiff EoS case, while maintaining a small but finite difference for the soft EoS. The observed energy evolution of $v_2$ for protons and deuterons offers crucial constraints on both the nuclear equation of state and light nuclei production mechanisms. The results could serve as valuable theoretical inputs for interpreting STAR measurements.

\section{Summary}
Using the JAM transport with nucleon coalescence model, we investigate the elliptic flow $v_2$ of protons and deuterons in mid-central ($b=4.3-8.5~{\rm fm}$) Au+Au collisions at $\ems=3.0-3.9$ GeV. Deuterons are formed via coalescence of nucleon pairs satisfying $\Dr < 4.5~{\rm fm}$ and $\Dp < 0.3~{\rm GeV}/c$.
For protons, the spectator squeeze-out effect generate negative $v_2$ at $\ems=$3.2 and 3.5 GeV, and accompanied by enhanced  $\langle\pt\rangle$ perpendicular to the reaction plane.
Notably, at $\ems = 3.2$ GeV, protons exhibit negative $v_2$ at mid-rapidity, while coalescence-produced deuterons show positive $v_2$, in agreement with STAR measurements.
To address the breaking of mass-number scaling in light nuclei $v_2$ at $\ems=3.0-3.9$ GeV, particularly the observed sign change at 3.2 GeV, we analyze the $dN/d\phi$ distributions of protons and deuterons in the rapidity window $0<y<0.5$. The results indicate that nucleons near $\phi=0$ or $\pi$ (along the reaction plane) are more likely to satisfy the coalescence criteria, leading to a higher coalescence probability compared to nucleons near $\phi=\pi/2$.
Additionally, the parameters $\Dr$ and $\Dp$ show a strong dependence on $\pt$, with larger values at higher $\pt$.
The enhancement of $\langle\pt\rangle$ near $\phi=\pi/2$ reduces the coalescence probability, which influences deuteron formation along the azimuthal angle and disrupts mass-number scaling in the energy range of several GeV. This effect, particularly, leads to the observed sign inversion of deuteron $v_2$ in mid-rapidity at 3.2 GeV. This provides deeper insights into the late-stage nucleon coalescence dynamics, particularly regarding  the mass-number scaling violation of light nuclei flow in heavy-ion collisions. It also helps enhance our understanding of NCQ scaling behavior at lower collision energies.

Furthermore, simulations with different values of nuclear incompressibility ($\kappa$ = 210 MeV and 380 MeV) show that the stiffness of the equation of state plays a critical role in the sign inversion of deuteron $v_2$ at $\ems=3.2$ GeV. This insight provides a valuable means to constrain the nuclear incompressibility parameter $\kappa$ using experimental data from STAR.

\section*{Acknowledgments}
We are grateful for discussions with Dr. Yasushi Nara. This work is supported in part by the National Key R\&D Program of China (No. 2024YFA1610700) and the National Natural Science Foundation of China under(No. 12205342).

\bibliographystyle{apsrev4-1}
\bibliography{ref}  

\end{document}